\documentclass[runningheads]{llncs}
\usepackage[T1]{fontenc}
\usepackage{booktabs}
\usepackage{amsmath}
\usepackage{multirow}
\usepackage{hyperref}
\usepackage{graphicx}
\begin{document}
\title{Foundation Model Fine-Tuning with Attention-based Multi-View Fusion for Multi-Task Cardiac Function Assessment}
\titlerunning{Foundation Model Fine-Tuning for Multi-Task Cardiac Function Assessment}
%
%
\author{João Freitas\inst{1, 2, 3}, 
Warre Snaet\inst{3, 4}, Jan D'hooge\inst{2} \and Sandro Queirós\inst{3}}  

\authorrunning{Freitas \textit{et al.}}
%
\institute{Life and Health Sciences Research Institute, University of Minho, Portugal \and
Cardiovascular Imaging and Dynamics Laboratory, KU Leuven, Belgium \and 2Ai - Applied Artificial Intelligence Laboratory, School of Technology, Polytechnic University of Cávado and Ave (IPCA), 4750-810 Barcelos, Portugal \and Howest University of Applied Sciences,
Kortrijk, Belgium}

\maketitle              
\begin{center}
\begin{minipage}{0.95\textwidth}
\footnotesize
\noindent
\textbf{Preprint notice:}
This manuscript corresponds to the original version submitted
to the MICCAI 2026 EchoRisk Challenge, prior to peer review.
The accepted version incorporates methodological revisions
and should be consulted for the final implementation.
The accepted manuscript is publicly available on the
official MICCAI 2026 Satellite Events website:
\href{https://papers.miccai.org/miccai-2026-sat/paper/EchoRisk2026_018.pdf}
{MICCAI 2026 Accepted Manuscript}.
\end{minipage}
\end{center}

\vspace{0.5em}

\begin{abstract}
In cardio‑oncology, left ventricular ejection fraction (LVEF) is the standard biomarker for monitoring cardiac function during cancer therapy. However, its estimation from echocardiography is highly operator dependent, and clinically meaningful declines are often detected only after myocardial injury has progressed, motivating automated methods for standardized assessment and earlier detection of LV dysfunction. To address this need, we propose a multi‑task framework integrating cardiac cycle-based temporal standardization, multi‑view spatiotemporal feature extraction, and joint optimization for LVEF estimation and LV dysfunction classification. Apical four‑ and two‑chamber echocardiographic sequences are temporally standardized by sampling a clip over an estimated cardiac cycle before being processed by the EchoPrime foundation model, which is fine-tuned end-to-end within the proposed architecture. View-specific features are fused through an attention‑based aggregation module to generate a patient‑level embedding, from which two task-specific prediction heads estimate continuous LVEF and binary LV dysfunction. To broaden the LVEF distribution encountered during training, samples from the MIMIC‑IV-ECHO dataset are incorporated using a balanced sampling strategy. Evaluated on the EchoRisk-MICCAI Challenge, our method achieves an MAE of 4.003 for LVEF estimation (Task 1) and an AUC of 0.764 for LV dysfunction detection (Task 2) on the validation set, demonstrating the effectiveness of foundation model fine-tuning with attention-based multi-view fusion for automated cardiac function assessment in cardio‑oncology.


\end{abstract}
\section{Introduction}
Echocardiography is the primary imaging modality for cardiac surveillance during cancer therapy, providing repeatable, non‑invasive assessment of cardiac structure and function without radiation exposure \cite{1}. In cardio‑oncology, left ventricular ejection fraction (LVEF) remains the standard biomarker for guiding follow‑up and treatment decisions. However, its estimation in routine clinical practice is highly operator dependent and subject to substantial inter‑observer variability, with clinically meaningful declines often detected only after myocardial injury has progressed \cite{2}. Consequently, LVEF is a relatively late marker of cardiotoxicity rather than a sensitive indicator of early cardiac dysfunction.

These limitations highlight the need for automated methods capable of estimating cardiac function while simultaneously identifying clinically meaningful deterioration. Such tools are particularly valuable for longitudinal patient monitoring, where consistent assessment across time, operators, and clinical centers is essential. Modern deep learning approaches are well suited to this challenge, as they can learn spatiotemporal representations directly from echocardiographic videos and capture subtle functional changes that may not be captured by conventional measurements. However, publicly available multicenter datasets for longitudinal cardiac function assessment in oncology populations remain scarce.

The EchoRisk dataset \cite{3} addresses this gap by providing longitudinal echocardiographic data collected across multiple centers. Building on this resource, the EchoRisk-MICCAI Challenge defines three clinically motivated tasks: LVEF estimation (Task 1), left ventricular (LV) dysfunction classification (Task 2), and early cardiotoxicity prediction from baseline imaging (Task 3). In this work, we focus on Tasks 1 and 2, both targeting LV systolic function. Their complementary objectives naturally motivate a multi‑task learning framework, where continuous LVEF regression provides detailed functional supervision while dysfunction classification encourages learning representations associated with clinically meaningful impairment \cite{3}. Accordingly, we propose a framework combining cardiac cycle-based temporal standardization, the EchoPrime foundation model with attention-based multi-view fusion, and joint optimization for LVEF estimation and LV dysfunction classification from apical two‑ and four‑chamber echocardiographic sequences.

\section{Methodology}

\subsection{Overview}
The proposed framework comprises several components (Fig. 1) that jointly capture cardiac-cycle dynamics while enabling multi‑view learning. First, temporal preprocessing standardizes each video to 16-frame clips sampled over an estimated cardiac cycle, inferred from LV segmentation and peak analysis of the area‑versus‑time curve. Second, the EchoPrime encoder \cite{4} extracts spatiotemporal features from each available view. These view-specific features are aggregated through an attention-based multi-view fusion module to generate a patient-level embedding, from which two task-specific prediction heads estimate continuous LVEF and binary LV dysfunction. Additional LVEF supervision is provided by samples from the MIMIC-IV-ECHO dataset \cite{5}. The entire framework is optimized end-to-end, allowing the pretrained encoder to adapt jointly with the attention module and prediction heads while leveraging the external LVEF supervision.

\begin{figure*}[!t]
    \centering
    \includegraphics[width=\textwidth]{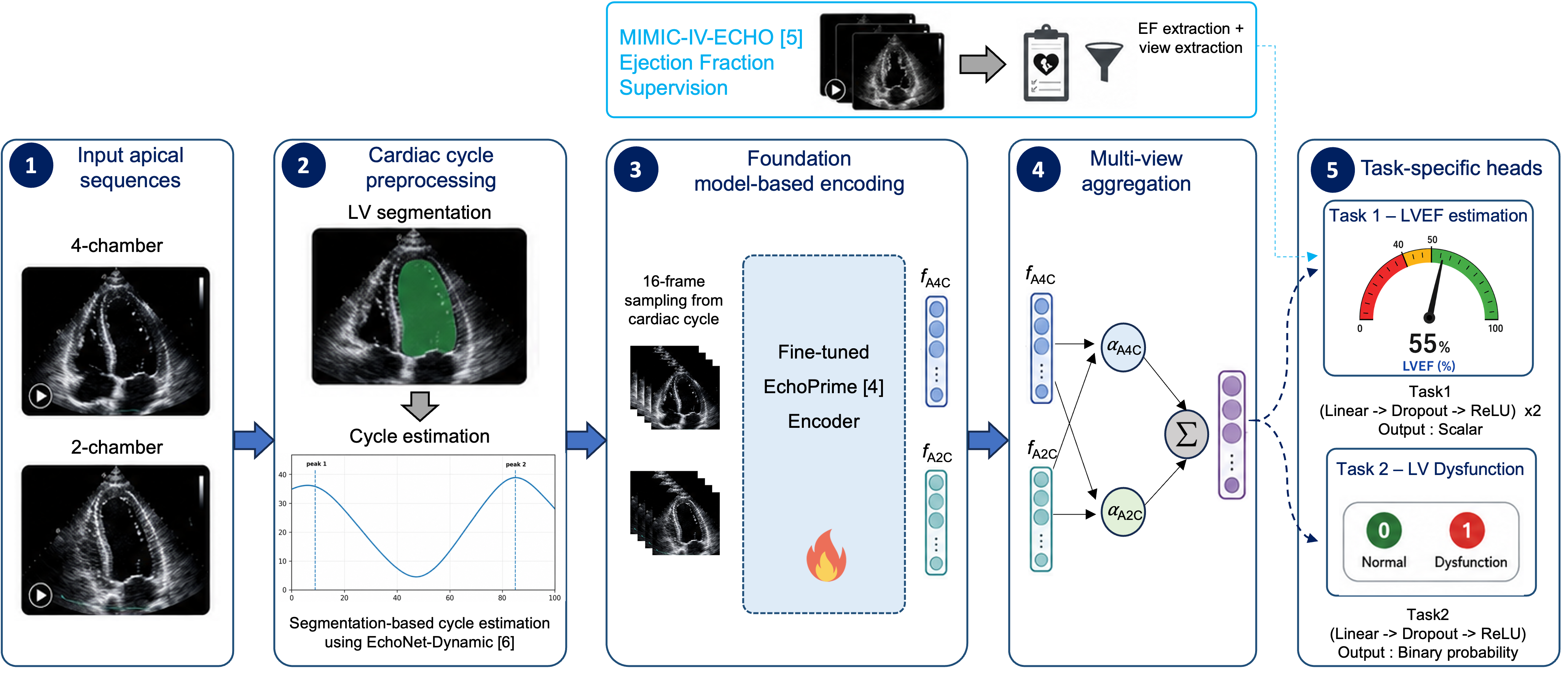}
    \caption{
    Overview of the proposed multi‑task framework.}
    \label{fig:framework_overview}
\end{figure*}

\subsection{Cardiac cycle video preprocessing}
Reliable temporal standardization is essential because both LVEF estimation and LV dysfunction detection depend on how cardiac motion is represented in the input video. The EchoRisk baseline \cite{3} addresses missing or unreliable DICOM frame‑rate metadata by applying fractional index sampling based on the video's frame count. While simple and robust, this approach implicitly assumes that uniformly sampling the entire video provides an adequate representation of cardiac motion, even when multiple cardiac cycles are present. To better reflect the underlying cardiac dynamics, we instead estimate the cardiac cycle duration directly from image content.

Specifically, we use the pretrained EchoNet‑Dynamic segmentation model \cite{6} to generate frame‑wise LV masks. The LV area is then computed for each frame, yielding an area-versus-time curve. Peak detection identifies end‑diastolic frames, from which the cardiac cycle duration is estimated. This duration subsequently guides uniform temporal sampling, producing standardized clips with consistent temporal coverage across patients.

\subsection{EchoPrime feature extraction}
EchoPrime  \cite{4} is a spatiotemporal foundation model designed for multi-view echocardiographic analysis. It was pre-trained on more than 12 million paired ultrasound videos and clinical reports, exposing the model to substantial variation in image quality, acquisition protocols, and cardiac morphology.

In our framework, each DICOM sequence is preprocessed using the original EchoPrime pipeline, including masking pixels outside the ultrasound field of view, followed by cropping and resizing the frames to 224×224 pixels. From the resulting video, temporally standardized 16-frame cardiac clips are extracted from each available apical view. EchoPrime combines convolutional feature extraction with transformer-based temporal modeling to produce a 512-dimensional embedding for each clip, summarizing cardiac structure and motion throughout the cardiac cycle.


\subsection{Multi-view aggregation}
Let each apical view $v \in \{\mathrm{A4C}, \mathrm{A2C}\}$ be encoded by the EchoPrime encoder, producing a feature embedding $F_v$. To provide the aggregation module with explicit view information, $F_v$ is concatenated with a one-hot encoding $e_v$ indicating the view identity, yielding the augmented representation $\tilde{F}_v$. This view-aware representation introduces a lightweight inductive bias that facilitates consistent feature interpretation across views while improving the stability of the attention mechanism. Attention logits are then computed as
\[
\ell_v = W_2 \, \tanh\!\left( W_1 \, \mathrm{LN}(\tilde{F}_v) \right),
\]
\noindent where $\mathrm{LN}(\cdot)$ denotes Layer Normalization, and $W_1$ and $W_2$ are learnable linear projections. Missing views are handled using a binary mask $m_v \in \{0,1\}$, ensuring that absent views receive zero probability under the Softmax applied to $\{\ell_v\}$.

The patient-level embedding is computed as

\[
h = \sum_{v} \alpha_v \tilde{F}_v,
\]
\noindent where $\alpha_v$ denotes the normalized attention weight assigned to view $v$. By combining view-aware representations with attention-based fusion, the model adaptively weights each view according to both its content and identity, improving robustness to missing or poor-quality views.

\subsection{Task-specific prediction heads}
The aggregated embedding $h$ is processed by two independent prediction heads, one for each downstream task. The LVEF head is implemented as a two-layer multilayer perceptron (MLP), where each layer consists of Layer Normalization, a linear projection, ReLU activation, and dropout, followed by a final linear layer for continuous LVEF prediction. The LV dysfunction head follows the same design but uses a single hidden layer before the final linear classifier. Both heads use a hidden dimensionality of 256 and a dropout rate of 0.4, providing sufficient capacity for task-specific adaptation while reducing overfitting.

\subsection{MIMIC-IV-ECHO Dataset Preparation}
Additional LVEF supervision is obtained from the MIMIC-IV-ECHO dataset \cite{5} through a dedicated preprocessing pipeline that automatically reads biplane LVEF measurements and identifies the corresponding view-specific videos from raw DICOM studies.

Each DICOM file is scanned to locate the measurement box using color‑based thresholding, leveraging the consistent annotation color across the dataset. If detected, the measurement region is cropped and binarized so that the text appears in white against a black background. Individual text lines are then segmented by computing the vertical projection profile and identifying contiguous foreground regions separated by background gaps along the y-axis. Each extracted line is subsequently processed with GLM‑OCR \cite{7} to recover the corresponding text. To improve extraction reliability, an additional validation step is performed using the horizontal projection profile. Similar to the line segmentation procedure, contiguous foreground regions separated by background gaps along the x-axis are identified to estimate the number of characters in each line. This count is compared with the number of characters returned by the OCR model, and inconsistent detections are discarded. Among the validated lines, we search for entries containing the text "EF Biplane", corresponding to LVEF measurements derived from both apical four- and two-chamber views.

For each study containing a valid measurement, apical four‑ and two‑chamber videos are identified using the pretrained EchoPrime view-classification head. Studies containing both views and a valid EF Biplane measurement are included as additional supervision for the LVEF regression task.

Of note, unlike the EchoRisk dataset, MIMIC-IV-ECHO provides temporal DICOM metadata that are used to directly estimate the cardiac cycle duration, avoiding the image-based procedure described in Section~2.2.

\subsection{Training strategy}
The model was trained for 50 epochs using the AdamW optimizer \cite{8} with a batch size of 8. All model parameters were optimized end-to-end, with learning rates of $1\times10^{-5}$ for the EchoPrime encoder and $2\times10^{-4}$ for the attention module and task-specific prediction heads, together with a weight decay of $1\times10^{-5}$. A warmup–cosine learning rate schedule \cite{9} was employed, consisting of a five-epoch warmup followed by cosine decay over the remaining training epochs. Early stopping with a patience of 10 epochs was performed independently within each cross-validation fold using the corresponding validation split. Training was terminated only when neither task improved within the patience window. The best-performing checkpoint for each task was saved separately, allowing the shared encoder to benefit from joint optimization while selecting the optimal model for each downstream task.

Data augmentation was applied on-the-fly to improve generalization. Spatial augmentations included random rotations of up to $10^\circ$ and translations of up to 8\% of the image size, while intensity augmentations comprised brightness and contrast jitter, Gaussian noise, and mild gamma adjustments. Each augmentation was applied with a probability of 0.5 and consistently across all frames within a clip to preserve temporal coherence. Training clips were generated using the temporal standardization procedure described in Section 2.2. During training, a random starting frame was selected for each clip, while the estimated cardiac cycle duration defined the temporal sampling interval, increasing temporal diversity and robustness to variations in clip alignment.

The network was optimized using a multi-task objective combining Smooth L1 regression for LVEF estimation ($\beta = 5.0$) and binary cross-entropy for LV dysfunction classification. To mitigate class imbalance, the latter incorporated positive-class weighting and a task weight of 0.5. LVEF targets were normalized to stabilize optimization. MIMIC-IV-ECHO samples contributed exclusively to the LVEF regression objective, with their loss weighted by 0.5 to provide auxiliary supervision while preserving the primary emphasis on the EchoRisk dataset. This strategy enables the model to benefit from the broader LVEF distribution available in MIMIC-IV-ECHO while maintaining specialization for the cardio-oncology population. The total loss is defined as

\[
\mathcal{L}_{\text{total}}
=
\left(
\mathcal{L}_{\mathrm{LVEF}}
+
0.5\,\mathcal{L}_{\mathrm{LVEF}}^{\mathrm{MIMIC}}
\right)
\;+\;
0.5\,\mathcal{L}_{\mathrm{LVDysfunction}}
\]

\section{Experiments and results}

\subsection{Dataset}
For Tasks 1 and 2, the EchoRisk dataset \cite{3} comprises 422 patients, 1,123 examinations, and 2,159 echocardiographic videos. The data are split at the patient level, with  approximately 56\%, 14\% and 30\% of the examinations assigned to the training, validation, and test sets, respectively. As the official test set is hidden during the challenge, the official validation set served as the final evaluation set during development. To avoid tuning model design decisions to this set, five patient-disjoint models were trained using cross-validation within the official training set. During evaluation, the predictions from all five models are averaged to obtain an ensemble prediction for each examination in the official validation set.

To broaden the LVEF distribution available during training, we additionally incorporate samples from the MIMIC-IV-ECHO dataset. Our preprocessing pipeline identified 3,278 examinations from 2,538 patients containing both apical four- and two-chamber views together with a valid biplane LVEF measurement. At each training epoch, a randomly selected subset of these examinations, matching the number of EchoRisk training examinations in the corresponding fold, is included to maintain a balanced contribution from both datasets and prevent the external data from dominating the optimization process.

\subsection{Overall impact of the proposed components}
To assess the contribution of each component, we evaluated different configurations of the proposed framework. Table~\ref{tab:full_experiments} reports performance on the official validation set using the ensemble predictions obtained from the five cross-validation models. Performance is measured using the mean absolute error (MAE) for LVEF regression (Task 1) and the area under the receiver operating characteristic curve (AUC) for LV dysfunction classification (Task 2). During validation, clips were extracted deterministically by centering the temporal sampling window within each video and uniformly sampling frames over the estimated cardiac cycle duration, providing a reproducible evaluation protocol while reducing the likelihood of incomplete cardiac cycles or probe repositioning artifacts.

\begin{table*}[!b]
\centering
\footnotesize
\setlength{\tabcolsep}{2pt}
\renewcommand{\arraystretch}{1}
\caption{Impact of the proposed components on Tasks 1 (MAE) and 2 (AUC), reported on the official validation set using the ensemble of the five cross-validation models. Best and second-best results for each metric are indicated in \textbf{bold} and \underline{underlined}, respectively.}  
\label{tab:full_experiments}

\begin{tabular}{l c c}
\toprule
\scriptsize
\textbf{Model} 
& \textbf{↓ MAE} 
& \textbf{AUC ↑} \\
\midrule

Original EchoPrime
& 6.970 & -- \\

Original EchoPrime + Cardiac-cycle standardization
& 5.770 & -- \\

\midrule

Frozen EchoPrime + Multi-task
& 4.314 & 0.677 \\

Fine-tuned EchoPrime + Multi-task
& 4.198 & 0.744 \\

Fine-tuned EchoPrime + Multi-task + MIMIC ($w = 1.0$)
& \underline{4.093} & 0.739 \\

Fine-tuned EchoPrime + Multi-task + MIMIC ($w = 0.5$)
& \textbf{4.003} & \underline{0.746} \\

Fine-tuned EchoPrime + Task 1 + MIMIC ($w = 0.5$)
& 4.152 & -- \\

Fine-tuned EchoPrime + Task 2
& -- & 0.719 \\

Fine-tuned EchoPrime + Multi-task + MIMIC ($w = 0.5$)
&  &  \\ 

(all training data, no early stopping)
& 4.213  &  \textbf{0.764} \\ 

\midrule

Challenge baseline 
& 5.020 & 0.755 \\

\bottomrule
\end{tabular}
\end{table*}

To isolate the effect of cardiac cycle-based temporal standardization, we first compared the original EchoPrime temporal sampling strategy with the proposed segmentation-guided sampling. Standardizing each input clip according to the estimated cardiac cycle duration substantially reduced the LVEF estimation error, demonstrating the importance of physiologically consistent temporal sampling.

Introducing the proposed multi-task framework with a frozen EchoPrime encoder further improved LVEF estimation, although LV dysfunction classification remained comparatively limited. This is consistent with the pretraining objective of EchoPrime, whose representations primarily capture global cardiac function. Fine-tuning the encoder on the EchoRisk dataset improved both tasks, highlighting the importance of adapting the pretrained representation to the target cardio-oncology domain.

Although MIMIC-IV-ECHO does not provide LV dysfunction labels, additional LVEF supervision may still influence the shared encoder and indirectly benefit the classification task. Incorporating MIMIC-IV-ECHO samples with equal loss weighting improved LVEF estimation but slightly reduced Task 2 performance, suggesting that excessive emphasis on the auxiliary regression objective biases the shared representation toward global systolic function. Reducing the MIMIC-IV-ECHO loss weight to 0.5 achieved the best trade-off, yielding the lowest MAE (4.003) and the second-highest AUC (0.746). These results indicate that external EF supervision is most effective as a complementary training signal rather than a dominant optimization objective.

The task-specific variants further support the value of multi-task learning, as both underperformed relative to the joint model. This suggests that the two objectives provide complementary supervision: Task 1 encourages learning a robust global representation of cardiac function, while Task 2 promotes sensitivity to clinically meaningful deterioration patterns.

Despite outperforming the Task 2-only variant, the joint model remained challenged by the limited number of positive samples available for learning LV dysfunction classification. This limitation is further amplified under five-fold cross-validation, where the already scarce positive cases are distributed across multiple training folds. To assess whether maximizing the available Task 2 supervision could further improve classification performance, we trained the best-performing configuration on the entire official training set without early stopping. This increased the AUC to 0.764, supporting the hypothesis that LV dysfunction detection benefits from exposure to additional positive training examples. In contrast, the MAE increased to 4.213, suggesting that this strategy was less beneficial for LVEF regression.

\section{Conclusion}
We presented a multi-task framework for automated cardiac function assessment in cardio-oncology, combining cardiac cycle-based temporal standardization, end-to-end adaptation of the EchoPrime foundation model, attention-based multi-view fusion, and auxiliary LVEF supervision from MIMIC-IV-ECHO. The proposed approach improved both LVEF estimation and LV dysfunction detection on the EchoRisk Challenge validation set, highlighting the potential of task-adapted foundation models for robust echocardiographic cardiac function assessment.

\begin{credits}

\subsubsection{\ackname} The authors acknowledge the financial support provided by Portuguese National funds, through the Foundation for Science and Technology (FCT) under grant 2024.03683.BD (J.F.).

\subsubsection{\discintname}
The authors declare that they have no competing interests.
\end{credits}
%
%
%
%

\end{document}